\documentclass[12pt,a4paper]{article}

\usepackage{amsbsy,latexsym,cite}
\usepackage{graphics}

\newcommand{\no}{\noindent}

\newcommand{\ga}{\gamma}
\newcommand{\de}{\delta}
\newcommand{\Ha}{{ H}_{\mathrm{int}}}
\newcommand{\psib}{\bar\psi}
\newcommand{\free}{{\mathrm{f}}}
\newcommand{\psibfree}{\bar\psi^\free}
\newcommand{\psifree}{\psi^\free}
\newcommand{\Afree}{A^\free}
\newcommand{\soft}{{\mathrm{soft}}}
\newcommand{\phase}{{\mathrm{phase}}}

\newcommand{\as}{{\mathrm{as}}}
\newcommand{\Haas}{{ H}_{\mathrm{int}}^{\as}}

\newcommand{\psias}{\psi^\as}
\newcommand{\Aas}{A^\as}
\newcommand{\bd}{{b^{\dag}}}
\newcommand{\dd}{{d^{\dag}}}
\newcommand{\ad}{{a^{\dag}}}
\newcommand{\cd}{\cdot}
\newcommand{\ecd}{{\cdot}}
\newcommand{\eps}{\epsilon}
\newcommand{\mod}[1]{\vert {#1}\vert}
\newcommand{\pa}{\partial}
\newcommand{\G}{{\cal G}}
\newcommand{\tfrac}[2]{{\textstyle\frac{#1}{#2}}}
\newcommand{\ee}{\mathrm{e}}
\newcommand{\ii}{{i}}

\newcommand{\psidirac}{\psi_D}
\newcommand{\psiF}{\psi_f}
\newcommand{\ket}[1]{\vert #1 \rangle}

\newcommand{\bra}[1]{\langle #1 \vert}
\newcommand{\normm}[1]{\frac1{2E_{#1}}}

\newcommand{\xb}{{\boldsymbol x}}
\newcommand{\yb}{{\boldsymbol y}}
\newcommand{\zb}{{\boldsymbol z}}
\newcommand{\pb}{{\boldsymbol p}}

\newcommand{\kb}{{\boldsymbol k}}
\newcommand{\vb}{{\boldsymbol v}}

\newcommand{\ab}{{\boldsymbol a}}
\newcommand{\intx}{\int\!d^3x\;}
\newcommand{\inty}{\int\!d^3y\;}
\newcommand{\intz}{\int\!d^3z\;}
\newcommand{\intp}{\int\!\frac{d^3p}{(2\pi)^3}\;}

\newcommand{\intk}{\int\!\frac{d^3k}{(2\pi)^3}\;}
\newcommand{\intks}{\int\limits_{\soft}\!\frac{d^3k}{(2\pi)^3}\;}

\newcommand\II{\textsl{II}}

\begin{document}

\begin{titlepage}
\rightline{BNL-HET-99/18}
\rightline{PLY-MS-99-23}

\vskip19truemm
\begin{center}{\Large{\textbf{Charges from Dressed Matter: Construction}}}\\ [8truemm]
 \textsc{Emili Bagan\footnote{email:
bagan@quark.phy.bnl.gov; permanent address: IFAE, Univ.~Aut\`onoma de
Barcelona,\\ E-08193 Bellaterra (Barcelona), Spain}\\[5truemm] \textit{Physics
Dept.\\ Brookhaven National Laboratory\\ Upton, NY 11973\\ USA}\\[10truemm]
Martin Lavelle\footnote{email: mlavelle@plymouth.ac.uk}, and
David McMullan}\footnote{email: dmcmullan@plymouth.ac.uk}\\
[5truemm] \textit{School of Mathematics and Statistics\\ The
University of Plymouth\\ Plymouth, PL4 8AA\\ UK} \end{center}

\bigskip\bigskip\bigskip
\begin{quote}
\textbf{Abstract:} There is a widespread belief in particle physics that there
is no relativistic description of a charged particle. This is claimed to be due to
persistent, long range interactions which distort the in and out going plane waves and
generate infra-red divergences.   In this paper we will
show that this is not the case in QED. We construct locally gauge invariant charged
fields which do create in and out Fock states. In a companion paper we
demonstrate
that the Green's functions of these fields have a good pole structure
describing particle propagation.
\end{quote}

\end{titlepage}

\setlength{\parskip}{1.5ex plus 0.5ex minus 0.5ex}



\section{Introduction}

The successes of hadronic physics are to a great extent based
on our use of partonic variables in the ultraviolet (UV)
domain. However, in the realm of soft physics, where the dynamics of
quarks and gluons underlie jet formation and
hadronisation, we simply do not know what variables to use. Strong
interaction phenomenology could be a primary source of information here
and this is, to a large extent, based upon effective quark models. However,
the \lq quarks\rq\ which compose these models are, although
undoubtedly useful in very many ways, in no sense organic to Quantum
Chromodynamics (QCD). This remains the case when such models are further
mutated in the hope of making them more closely resemble
the low energy dynamics of QCD, e.g.,
by introducing \lq gluons\rq\ into effective quark models.
Whatever these effective variables are, they are not simply the
Lagrangian fields of QCD.

Knowledge of the correct infra-red (IR) variables in
QCD would help us to obtain a deeper
understanding of the structure of the underlying field
theories of the standard model, to obtain some feeling for the validity and
the ({\it a priori} unknown) limits of phenomenological models in
particle physics and, most importantly, help us understand
hadronisation and confinement. This paper is devoted to the
construction of physical charged fields
using only the gauge theories of the fundamental interactions. Faced
with such a major task, we cannot hope to solve it immediately and our
procedure will be essentially perturbative. Our aim is to develop
methods and insight which may be carried over and extended into the
non-perturbative domain. We intend to show below that this aim is not
completely unrealistic.

Let us first recall how the problems associated with the use of
unphysical variables show themselves in the standard description of
scattering. In the LSZ approach it is assumed that at
times long before or after any scattering process the fields
entering or emerging from the vertex do not interact with each other
any more. The experimental fact of hadronisation tells us that this
\emph{cannot} be true if our asymptotic fields are just taken to be the
Lagrangian quark fields. Although we can describe the
hadronisation process in one of the many models, this is necessarily matched on to
partons emerging from the scattering process. This has even been
elevated to the status of the principle of local parton hadron
duality~\cite{Azimov:1985np,Azimov:1986by}. However, this matching
onto free partons has a price:
quarks and gluons still interact with each other even when widely
separated. The various  IR~divergences
in on-shell Green´s functions and in $S$-matrix elements
are a direct consequence of the neglect of such interactions. In fact even
in Quantum Electrodynamics (QED),
where there is a $1/r$ fall-off in the interaction between
the charges which we will generically refer to as electrons, it has
long been known~\cite{Dollard:1964,Chung:1965,Kibble:1968a,Kibble:1968b,Kibble:1968c,
kulish:1970,havemann:1985,Horan:1998im}
that this fall-off is still too slow to permit us to ignore the interaction
and its neglect is responsible for the IR~catastrophe.

There are various responses to this situation: the one most
widely, if tacitly, employed
has been to give up any description of on-shell Green´s
functions and $S$-matrix elements and to restrict oneself to IR safe
quantities, such as, in QED, cross-sections summed over all final state
soft photons. This response is at once theoretically radical (discarding
the $S$-matrix) and practically
conservative (we should only attempt to talk about this class of measurable
quantities). It may be noted that this posture does not address the
questions initially raised above, viz.\ what are the correct IR degrees
of freedom? A second approach is to redefine
the division of the Hamiltonian into free and interaction terms so
that the new interaction can indeed be neglected
at asymptotic times. This approach has been studied for QED by
various authors \cite{Dollard:1964,kulish:1970,Zwanziger:1975jz,Zwanziger:1975ka}
but has not been successfully extended to QCD.
As stressed by the authors of \cite{kulish:1970}, taking this route
entails the loss of a particle description already at the level of
QED! (The perturbative signal for this is that the
on-shell QED Green's functions so obtained, although IR~finite, do not
have a pole structure describing particle propagation, rather the external lines
for electrons and other charged fields have branch functions.)
A third reaction, which we pursue below, is
to argue that the experimentally observed
particles entering and leaving scattering processes do not
directly correspond to the variables which we have in the Lagrangian.
E.g., when an electron emerges from a scattering process it is
accompanied by an electromagnetic field which is not accounted for by
the free Lagrangian fermion, $\psi$. We will refer to the inclusion
of such fields as \emph{dressing} the Lagrangian matter field.
Such ideas have been investigated by many authors, see,
e.g.,~\cite{Dirac:1955ca,Steinmann:1984,dEmilio:1984,
dEmilio:1984a, Steinmann:1984, Prokhorov:1992, Prokhorov:1993,
Kawai:1995zx,Kawai:1995zv, Kawai:1995zu, Lusanna:1996ut,
Lusanna:1996us, Lavelle:1997ty, Kashiwa:1997,
Kashiwa:1996, Chechelashvili:1997, Haagensen:1997pi}.

We will argue below that this last approach, choosing the correct
asymptotic fields, indeed solves the problem in a way which is
compatible with having the
expected pole structure, retaining the $S$-matrix and having a particle
description. Here of course we need to understand what are
the \lq correct\rq\ fields. We will
present below two requirements for the construction of
such fields and further
show how the descriptions which satisfy these conditions
may be obtained in a \emph{systematic fashion} which can be extended to
other theories and which will permit the eventual inclusion of
non-perturbative dynamics.

We have seen that the interaction in gauge
theories cannot be neglected at large times. Now the
Lagrangian fields in  QED transform under gauge transformations as
\begin{equation}\label{gauge}
 A_\mu(x)\to A_\mu(x) + \partial_\mu\theta(x)\,,\qquad
 \mathrm{and}
 \qquad\psi(x)\to \ee^{ie\theta(x)}\psi(x)
  \,,
\end{equation}
so, since the coupling does not vanish, the matter field,
$\psi$, is \emph{not} gauge invariant at large times and cannot be
identified with a physical particle such as an electron even in
the asymptotic region. Can we then construct a gauge invariant charged
field at all?

To make this more precise, we can use the tools of constrained
dynamics (see, e.g., \cite{Henneaux:1992ig}). The prime
requirement, as far as we are concerned in this paper, is that a
physical field must satisfy Gauss' law, which reads
\begin{equation}\label{gausslaw}
  \partial^i F_{i0}=-e J_0
\,.
\end{equation}
Since Gauss' law generates gauge transformations,
this translates into the requirement of
gauge invariance. There is a great deal of work in the
literature on gauge invariant variables or simply variables with
simplified gauge transformation properties, see, e.g.,
\cite{Mandelstam:1968hz,Haagensen:1996py,Gogilidze:1998qq}.
However, the identification of such variables with charged particles
is especially subtle. The main point we wish to note here is that
Gauss' law tells us that charged matter fields cannot be separated from
the electromagnetic cloud which surrounds them.
This implies then that any description of a physical charged particle
cannot be local since the total charge can always
be written as a surface integral at spatial infinity~\cite{Lavelle:1997ty}.
We should also point out that, since the charge density generates
global (rigid) gauge transformations, charged particles are not
invariant under such gauge trans\-format\-ions.

The first construction of charged fields that we are aware of was
performed by
Dirac~\cite{Dirac:1955ca}, who noted the existence of a set of
composite fields
\begin{equation}\label{fdress}
   \psiF(x)\equiv \exp\left( -ie\int\!d^4zf^\mu(x-z)A_\mu(z)
   \right)\psi(x)\,,
\end{equation}
which, Dirac argued,  are locally (but not globally) gauge
invariant for any $f^\mu$ so long as
$\pa_\mu f^\mu(w)=\delta^{(4)}(w)$ holds.

It is easy to see that there are many gauge invariant descriptions (many functions $f$
which fulfil this property). One of the main ideas behind the work reported here
is that gauge invariance is not enough: a physical field must single
itself out by further requirements. It is easy to see that
some constructions are gauge invariant but highly unphysical:
the most obvious example being the stringy ansatz
\begin{equation}\label{string}
   \psi_{\Gamma}(x)\equiv \exp\left( -ie\int_\Gamma^x dz_i
   A_i(x_0,\boldsymbol{z})
   \right)\psi(x)\,,
\end{equation}
which, it can be shown~\cite{Haagensen:1997pi},
corresponds to an infinitely excited state
where the electric flux is confined along the path of the
string, $\Gamma$.  The reader can find  an animation showing
the decay of such a state
at {\tt http://www.ifae.es/\~{}roy/qed.html}.

Dirac suggested the requirement
that, out of the general set of functions given by Eq.~\ref{fdress},
the physical description should have the correct electric field.
Concretely, he noted that a physical charged particle is
accompanied by an electric field and argued that an electron should
be described by
\begin{equation}\label{diracelec}
 \psidirac(x)\equiv \exp\left( -ie\frac{\partial_i
 A_i}{\nabla^2}\right)\psi(x)
\,,
\end{equation}
which is easily seen to be a special case of~(\ref{fdress}). His
argument for this choice was that the state, $\psidirac(x)\ket{0}$,
has a Coulombic electric field
\begin{equation}\label{stt}
 E_i(x_0,{\boldsymbol{x}})\psidirac(y)\ket0=
-\frac {e}{4\pi}
\frac{{\boldsymbol{x}_i-\boldsymbol{y}_i}}{\vert{\boldsymbol{x}-
\boldsymbol{y}}\vert^3}\psidirac(y)\ket0\,.
\end{equation}
To obtain this result\footnote{More generally, e.g., in Coulomb gauge where the
exponential factor reduces to unity, one needs to employ Dirac brackets.},
he used the  equal-time commutator:
$[E_{i}(x),A_{j}(y)]=\ii \delta_{ij}\delta(\xb-\yb)$. We note that
$\psidirac$ is locally gauge invariant and
that the factor of $1/\nabla^2$ in (\ref{diracelec})
shows the non-locality of the electromagnetic cloud around the
charge, since
\begin{equation}\label{sttt}
\frac1{\nabla^2}f(\xb)=-
\frac1{4\pi}\int\!d^3x\frac{f(\yb)}{\mod{\xb-\yb}}
\,.
\end{equation}
Since no spatial direction is singled
out, we might naively expect this to probably describe a static
charge and this is in agreement with (\ref{stt}).

A natural question now is how we should interpret this gauge invariant
field? Dirac argued that it corresponds to a charge at $\yb$ together
with its associated Coulombic electric field. Since the Coulomb field
is that of a static charge, his interpretation implies that we know
both the position \emph{and} the velocity of the charge! This,
as is well known from heavy quark effective theory (HQET), is
only compatible with the uncertainty principle in the infinite mass
limit.
This will be clarified in Sect.~3 for charges with finite masses.

An obvious test
of Dirac's construction is to study whether these fields
remove the IR singularities associated with on-shell static charges.
The simplest test is the two-point function of charges described by
(\ref{diracelec}) in an on-shell renormalisation scheme. It
was shown in~\cite{Lavelle:1993wy}
that this is indeed IR finite so long as the \emph{static}
point on the mass shell is chosen, i.e., $p=(m,0,0,0)$. (The
generalisation of this to the propagator at an arbitrary
point on the mass shell was shown in \cite{Bagan:1997dh}
for scalar QED and in \cite{Bagan:1997su} for the fermionic
theory.)

We note that it was
further shown in \cite{Lavelle:1994xa} how to extend
this description of dressed electrons to construct gauge invariant
dressed quarks in non-abelian gauge theories at any order in
perturbation theory (see the
appendix of \cite{Lavelle:1997ty}). Essentially  there is a minimal
perturbative extension of the QED result which retains gauge invariance.
We note that Haller and his collaborators \cite{Chen:1998cc,Haller:1997}
have also obtained perturbative formulae which agree with this
minimal extension of the static charge (\ref{diracelec}).

Three important implications of this work for QCD are that
colour charges are only well defined~\cite{Lavelle:1996tz} for locally
gauge invariant fields such as our dressed quarks and gluons; the identification
of a topological obstruction to the construction of an isolated
quark~\cite{Lavelle:1997ty} and the perturbative
identification~\cite{Lavelle:1998dv} of the dominant gluonic
configuration responsible for asymptotic freedom.

In the rest of this paper we will present a method to construct
charged particles in QED. We will systematically obtain explicit
solutions  and further demonstrate that the charged fields we
construct, since they are surrounded by the correct electromagnetic
fields, obey a free asymptotic dynamics. It will become clear that
our results reflect previous work on the structure of the asymptotic
dynamics which is found when it is assumed that the fields entering
and leaving scattering processes are just the free matter fields, however,
in our approach we will retain a particle interpretation.

In a companion paper~\cite{Bagan:1999yy}, \II, the results which we obtain below,
and the interpretation which we give to them, will be
further verified. In particular, we will use our physical variables
and study the IR behaviour and UV properties of various Green's
functions and $S$-matrix elements.

The structure of this paper is then as follows. In Sect.~2 we discuss
the asymptotic fields and dynamics of QED in some detail. Then, in
Sect.~3, we show how to correctly characterise charges and construct
static charges. The solution  for a charge in QED with an arbitrary,
relativistic velocity is presented in Sect.~4. In Sect.~5 the field
theoretical implications of this approach are briefly reviewed and in
Sect.~6 we present some conclusions and discuss the implications of
this paper.


\section{Dynamics and Asymptotic Dynamics}

The starting point for our analysis of the physical, charged sector of QED is
the familiar Lagrangian density
\begin{equation}\label{2qedlag}
{\cal
L}=-\tfrac14F_{\mu\nu}F^{\mu\nu}+i\psib\ga^\mu(\pa_\mu-ieA_\mu)\psi-m\psib\psi
+\tfrac12B^2+B\pa_\mu A^\mu \,.
\end{equation}
Note that here we use Feynman gauge, $B=-\pa_\mu A^\mu$, for a discussion of other
gauges, see, e.g.,~\cite{Haller:1993uf}. There is a significant algebraic advantage
to working in this gauge which leads to a much simpler account of
our construction of charges. We insist, though, that our
fundamental variables (which will be introduced in the next
section) are themselves gauge invariant and hence insensitive to
this choice of the Feynman gauge. In the companion paper \II, we will
indeed see the gauge invariance of our construction in many
perturbative calculations.

The extraction of the equal-time commutation relations from (\ref{2qedlag}) is,
at least, formally straightforward even in a fully interacting theory. What is
much more difficult to do is to construct the general space-time commutators of
these Heisenberg fields. For the $B$ field, though, things are much better
since in Feynman gauge it satisfies the free equations of motion ($\Box B=0$).
This fact, in conjunction with the known equal-time commutation relations,
leads to the space-time commutators \cite{nakanishi:1990}:
\begin{equation}\label{2bb}
  [B(x),B(y)]=0\,,
\end{equation}
\begin{equation}\label{2ba}
  [B(x), A_\mu(y)]=i\pa_\mu^xD(x-y)
\end{equation}
and
\begin{equation}\label{2bp}
  [B(x), \psi(y)]=e D(x-y)\psi(y)\,.
\end{equation}
where
{\setlength\arraycolsep{2pt}
\begin{eqnarray}\label{2d}
  D(x-y)&=&-\frac1{2\pi}\eps(x^0-y^0)\de((x-y)^2)\\
  &=&-\intk\frac1{\omega_k}e^{i\kb\ecd(\xb-\yb)}\sin
  \Bigl(\omega_k(x^0-y^0)\Bigr)\,,
\end{eqnarray}}%
and  $\omega_k=|\kb|$.
We stress that these commutators are valid for all times and, in particular,
hold both  at very early and at very late times.

The significance of these commutation relations comes from the role of the $B$
field in characterizing the physical states and observables of the theory.
Given its trivial dynamics, we can expand this field in terms of its modes:
\begin{equation}\label{2bc}
  B(x)=\intk\frac1{2\omega_k}\Bigl\{c(k)\ee^{-ik\ecd x}+c^\dag(k)\ee^{ik\ecd
  x}\Bigr\}\,.
\end{equation}
The condition on physical states
 is then that they are annihilated by the positive
frequency part\footnote{The connection between  this, by construction the positive frequency part of
$B$, and the gauge and matter fields will be clarified in Eq.~(37).
} of $B$, i.e.,
\begin{equation}\label{2phys}
  c(k)\ket{\mathrm{phys}}=0\,,
\end{equation}
for all $k$. Observables must preserve this condition, and hence must commute
with $c(k)$. From the space-time commutators (\ref{2ba}) and (\ref{2bp}) we
find that
\begin{equation}\label{2ca}
  [c(k),A_\mu(y)]=-ik_\mu\ee^{ik\ecd y}\,
\end{equation}
and
\begin{equation}\label{2cp}
  [c(k),\psi(y)]=-ie\,\ee^{ik\ecd y}\psi(y)\,.
\end{equation}
The first of these relations expresses the well known fact that
not all components of the vector potential are physical: at the
end of the day there should be just two  physical photonic degrees
of freedom per space-time point.
In contrast, the second relation
(\ref{2cp}) is not usually interpreted as saying that the matter
fields are  unphysical. This is because this commutator contains
an explicit dependence on the coupling constant. The widespread assumption
then is that, in the remote past and future, the interaction
vanishes and hence the asymptotic matter fields can be taken as
physical. There is, though, no obvious reason why the large time
limit of (\ref{2cp}) should vanish. Indeed, the long range nature
of the electromagnetic interaction suggests that it is highly
unlikely that a non-interacting regime is ever reached. In fact
these arguments are usually expressed \cite{itzykson:1980} as an
{\it ad hoc} assumption that the interaction is
adiabatically switched off in the remote past and future.

The appeal to such  a \lq switching off\rq\  mechanism for the
interaction does not reflect the experimental set-up found in
a typical scattering process. Rather, what happens is
that the in-coming, or out-going, particles become widely
separated and that it is this growing separation that effectively
renders the coupling to vanish. In theories with no more than
cubic interactions (such as spinorial QED) there is a simple
argument first used (to the best of our knowledge) by
Kulish and Faddeev \cite{kulish:1970} to indicate that in theories with
massless fields, this assumption is not true, i.e., the asymptotic
dynamics in QED is not that of the free theory.

The idea is very simple\footnote{Although to make this argument more rigorous  and
applicable to a wider class of theories a bit more care is needed. In particular we
stress that all such limits should be understood as {\it weak} limits between appropriate,
normalisable states. The full
construction with extensive applications will be reported
elsewhere~\cite{Horan:1999as}.}: go into
the interaction picture and take the large time limit of the interaction. If
this limit is zero, then the asymptotic dynamics is that of a free theory. If
the interaction does not vanish, then we simply cannot take the asymptotic
dynamics to be free.

In QED this works as follows. The interaction Hamiltonian in the interaction picture is given by
\begin{equation}\label{2intham}
  \Ha(t)=-e\intx \Afree_\mu(t,\xb)J^{\free\,\mu}(t,\xb)\,,
\end{equation}
where the free matter current is
$J^{\free\,\mu}(t,\xb)=\psibfree(t,\xb)\gamma^\mu\psifree(t,\xb)$
and we are using the superscript $\free$ to distinguish the free
fields from the corresponding interacting or Heisenberg  ones. To further fix our
notation, we take as our free field expansions
\begin{equation}\label{2psifree}
\psifree(x)=\intp\frac1{\sqrt{2E_{\smash{p}}}}\left\{
b(p,s)u^s(p)e^{-ip\ecd x}+
\dd(p,s)v^s(p)e^{ip\ecd x}
\right\} \,,
\end{equation}
where $E_p=\sqrt{|\pb|^2+m^2}$, a sum over $s$ is understood  and
\begin{equation}\label{2afree}
\Afree_\mu(x)=\intk\frac1{2\omega_{\smash{k}}}\left\{
a_\mu(k)e^{-ik\ecd x}+
\ad_\mu(k)e^{ik\ecd x}
\right\} \,.
\end{equation}
 We note here
the basic space-time commutator for the \emph{free} vector potential
\begin{equation}\label{2acomm}
  [\Afree_\mu(x),\Afree_\nu(y)]=-ig_{\mu\nu}D(x-y)\,.
\end{equation}

Inserting these free field expansions directly into
(\ref{2intham})  results in eight terms which can be  grouped
according to the positive and negative frequency components of the
\hbox{fields}. Each of these pieces will have a time dependence of
the form $e^{i\psi t}$ where $\psi$ involves sums and differences of
energy terms. The argument used in \cite{kulish:1970} (see also
\cite{Horan:1998im} and the discussion in
supplement S4 of \cite{jauch:1976}) is that, as $t$ tends to plus or minus infinity, only
terms with $\psi$ tending to zero can survive and thus contribute
to the asymptotic interaction. After performing the spatial
integration, and using the resulting momentum delta function, one
finds that in a theory describing massive charges only four terms of
the form $\psi=\pm (E_{p+k}-E_p\pm\omega_k)$ can survive in the large
time limit. The requirement that
$E_{p+k}-E_p\pm\omega_k\approx0$ can \emph{only} be met in QED
because the photon is massless, in which case it implies that
$\omega_k\approx0$, i.e., only the infra-red regime contributes to
the asymptotic dynamics. From this observation it can be shown
\cite{kulish:1970,Horan:1999as}  that the full interacting Hamiltonian
(\ref{2intham}) does \emph{not} vanish asymptotically\footnote{If the
photon is given a small mass, as an infra-red regulator, then the asymptotic Hamiltonian does
vanish.}, but has in fact the same asymptotic limit as
\begin{equation}\label{2hintas}
\Ha^{\as}(t)=-e\intx
\Afree_\mu(t,\xb)J^\mu_{\as}(t,\xb)
\end{equation}
with
\begin{equation}\label{2jas}
  J^\mu_{\as}(t,\xb)=\intp\frac{p^\mu}{E_p}\rho(p)\delta^3
  \bigl(\xb-t\pb/E_p\bigr)\,.
\end{equation}
The operators in this current are only contained in the
charge density
\begin{equation}\label{2cd}
  \rho(p)=\sum_{s}\Bigl(\bd(p,s)b(p,s)-\dd(p,s)d(p,s)\Bigr)
\end{equation}
which implies that the asymptotic current satisfies the trivial space-time
commutator relation
\begin{equation}\label{2jcomm}
[J^\mu_{\mathrm{as}}(x),J^\nu_{\mathrm{as}}(y)]=0\,.
\end{equation}
As such, this asymptotic current can be interpreted as effectively  the
integral over all momenta of the current associated with a charged
particle moving with velocity $\pb/E_p$. This does not
vanish as $t\to\infty$. We thus see that \emph{the asymptotic
dynamics of QED is not that of a free theory.}

\begin{figure}[htbp]
\begin{center}
\includegraphics{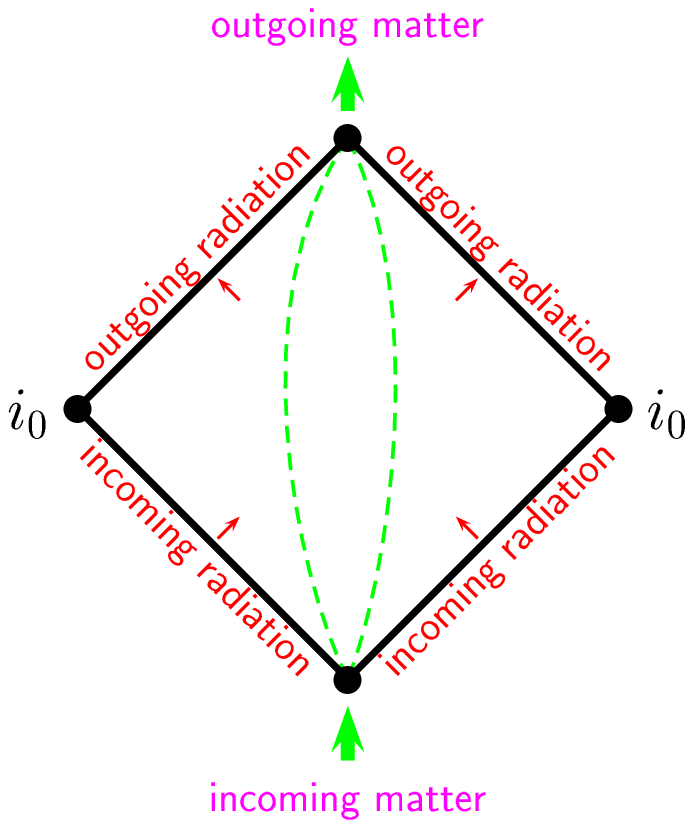}
\end{center}
\caption{Penrose diagram of Minkowski space-time:
the dashed lines are the worldlines of
massive particles while the short arrows depict incoming and outgoing
radiation. We may only demand that fields vanish at space-like
infinity, $i_0$.}
        \label{prdg}
\end{figure}

It is important to note here that by \lq asymptotic dynamics\rq\ we are referring to the dynamics in the neighbourhood
of time-like and null boundaries of Minkowski space-time. At space-like infinity things are very different: the
fields vanish and we impose the condition that the local gauge transformations
become the identity there. This distinction is made clear in the  Penrose diagram \cite{penrose:1972}
of Figure~\ref{prdg}.

What this argument has shown us is that, contrary to the statements found in
the usual approaches to scattering in QED, the basic commutator (\ref{2bp})
really does imply that the matter field is not physical even in the asymptotic
regime. \emph{The Lagrangian fermion in QED should not be associated with anything physical
such as an electron!} So what can we identify with physical matter?

Given that we are dealing with a gauge theory with an unbroken
symmetry, so that the reduction in the degrees of freedom must be
 in the gauge sector of the theory, it is clear \cite{Lavelle:1995rh} that we
cannot now  try to reduce  the  degrees of freedom in the matter sector.
Rather, a procedure is needed for finding
a gauge invariant combination of both  matter and gauge fields ---
we call this general process \emph{dressing the matter fields}.
This will be the topic of the next section.
Before moving on to that construction, though, it is useful to
continue our previous analysis of the asymptotic form  of the
matter and gauge fields.

It is important\footnote{In \cite{kulish:1970} and \cite{jauch:1976} the form
of the asymptotic fields were found by just translating from the interaction to
the Heisenberg picture, this resulted in a rather \emph{ad hoc} argument for why the
contribution from the lower  time $t_0$  could be ignored in the Heisenberg
fields.} to realise that there are two clear steps in the extraction of the
asymptotic form of the fields entering into our theory.
\begin{itemize} \item First we go
from the Heisenberg fields to the interacting fields via the usual time ordered
exponential of the interacting Hamiltonian:
\begin{equation}
O_{\mathrm{I}}(t,t_0)= T\exp\left(-i\!\!\int_{t_0}^t\!d\tau\,\Ha(\tau)\right)
O_{\mathrm{H}}(t)\,\widetilde{T}\exp\left(i\!\!\int_{t_0}^t\!d\tau\,\Ha(\tau)\right)\,,
\end{equation}
where $T$ and $\widetilde{T}$ signify time and anti-time ordering respectively.
This transformation requires the
introduction of an arbitrary time $t_0$ at which the fields in the two pictures
agree. This time is arbitrary at this stage because if we were to directly
transform back to the Heisenberg fields all reference to $t_0$ would vanish. At
lowest order in the coupling we find that the  relation between a Heisenberg
operator $O_{\mathrm{H}}(t)$ and its interaction picture counterpart is
\begin{equation}
O_{\mathrm{I}}(t,
t_0)=O_{\mathrm{H}}(t)-i\int_{t_0}^td\tau\,[\Ha(\tau),O_{\mathrm{H}}(t)]\,.
\end{equation}
  \item The asymptotic dynamics for the Heisenberg fields is
  now found by using the asymptotic form of the interaction
  Hamiltonian to translate from the interaction to the
  Heisenberg picture. At lowest order in the coupling we get
{\setlength\arraycolsep{2pt}
\begin{eqnarray}
O_{\mathrm{H}}^\as(t, t_0)&=&O_{\mathrm{I}}(t,
t_0)+i\int_{t_0}^td\tau\,[\Ha^\as(\tau),O_{\mathrm{I}}(t,t_0)]\nonumber\\
&=&O_{\mathrm{H}}(t)+i\int_{t_0}^td\tau\,[\Ha^\as(\tau)-\Ha(\tau),O_{\mathrm{H}}(t)]\,.
\end{eqnarray}}%
\end{itemize}
Note that there is a  $t_0$ dependence in this expression for the
asymptotic fields at time $t$ due to the potential contribution from the lower
range of integration. This dependence is governed by the value of
$\Ha^\as(t_0)-\Ha(t_0)$, so if we set $t_0=\pm\infty$ this will vanish by the
construction of the asymptotic interaction.

In practice, though, we do not want to start in the Heisenberg picture, rather
we wish to go directly from the known fields in the interaction picture. What
this argument then tells us is that we are free to ignore any explicit
contributions from the overlap time $t_0$ as long as this is taken to be
large.

The transformation from the  interaction picture to the asymptotic Heisenberg
picture is implemented by the usual time ordered product of the Hamiltonian
(\ref{2hintas}). Using the commutator relations (\ref{2jcomm}) and (\ref{2acomm}),
this can be written  as the product of two commuting terms:
\begin{equation}\label{2trans}
  T\exp\!\!\left(i\!\!\int_{-\infty}^t\!\!\!\!\!\!d\tau\,\Haas(\tau)\right)=
  \exp\!\!\left(i\!\!\int_{-\infty}^t\!\!\!\!\!\!d\tau\,\Haas(\tau)\right)
  \exp\!\!\left(\tfrac12\!\!\int_{-\infty}^t\!\!\!\!\!\!d\tau_1
  \int_{-\infty}^{\tau_1}\!\!\!\!\!\!d\tau_2\,[\Haas(\tau_1),\Haas(\tau_2)]\right)\,.
\end{equation}
From (\ref{2jcomm}) and (\ref{2acomm}), the
double commutator $[[\Haas(\tau_1),\Haas(\tau_2)],
\Afree(x)]=0$, and hence the asymptotic vector potential is given by ($t=x^0$)
{\setlength\arraycolsep{2pt}
\begin{eqnarray}\label{2aas}
 \Aas_\mu(x)&=&\exp\!\!\left(i\!\!\int_{-\infty}^t\!\!\!\!\!\!
 d\tau\,\Haas(\tau)\right)\Afree_\mu(x)
 \exp\!\!\left(-i\!\!\int_{-\infty}^t\!\!\!\!\!\!d\tau\,
 \Haas(\tau)\right)\nonumber\\
&=&\Afree_\mu(x)-e\int_{-\infty}^t\!\!d\tau d^3y\,D(\tau-t,
\yb-\xb)J^\as_\mu(\tau,\yb)\,.
\end{eqnarray}}%
Note that it follows from this that
\begin{equation}\label{2boxa}
 \Box\Aas_\mu(x)=-eJ^\as_\mu(x)\,,
\end{equation}
and hence we see that the asymptotic vector potential is made from  a free part plus
the field generated by the non-trivial asymptotic current, i.e., a response to
the presence of any other charges. We also have from (\ref{2acomm})
and (\ref{2jcomm}) that the asymptotic fields obey the same
commutator as the free fields
\begin{equation}\label{2afcomm}
  [\Aas_\mu(x),\Aas_\nu(y)]=-ig_{\mu\nu}D(x-y)\,.
\end{equation}

The asymptotic form of the  matter field follows in much the same way,
but now both terms in (\ref{2trans})
contribute. The resulting field is \cite{kulish:1970}
\begin{equation}\label{2psias}
\psias(x)=\intp\frac1{\sqrt{2E_{\smash{p}}}}D(p,t)
\left\{
b(p,s)u^s(p)e^{-ip\ecd x}+
\dd(p,s)v^s(p)e^{ip\ecd x}
\right\}
\end{equation}
where the distortion operator $D(p,t)$ factors into two commuting terms
(reflecting the similar factorisation in
(\ref{2trans})): $D(p,t)=D_{\phase}(p,t)D_{\soft}(p,t)$.
The phase component $D_{\phase}(p,t)$ does not depend on the gauge
field (it is in fact gauge invariant) and only contributes to the phase of the operator
(see the discussion in Supplement S4 of
\cite{jauch:1976}), as such we will neglect
its contribution to the asymptotic field. We do note, though, that this phase depends on
the presence of other charges in the system and hence enters into the
distortion operator only at order $e^2$ and above in the coupling.

The soft component of
the distortion operator is given by
\begin{equation}\label{2dsoft}
  D_{\soft}(p,t)=\exp\left\{-e\!\!\!\intks\frac1{2\omega_k}
  \left(
  \frac{p\cd a(k)}{p\cd k}\ee^{-itk\ecd p/E_p}-
\frac{p\cd \ad(k)}{p\cd k}\ee^{itk\ecd p/E_p}
  \right)\right\}\,,
\end{equation}
where we have made explicit that  only the soft
(infra-red) regime  contributes to the asymptotic fields.
Due to  this distortion operator, the asymptotic field $\psias$ does not create
or annihilate particle states. That is, if  we were to define
(as we would want to in the LSZ-reduction of the $S$-matrix) the
operator
\begin{equation}\label{2bbad}
  b(q,s, t):=\intx\frac1{\sqrt{2E_{\smash{q}}}}u^{\dag s}(q)\psias(x)\ee^{iq\ecd
  x}\,,
\end{equation}
then rather than recovering the time independent, one particle annihilation
operator, we get (neglecting the phase)
\begin{equation}\label{2bbadr}
  b(q,s,t)=D_\soft(q,t)b(q,s)\,.
\end{equation}
This operator does not have a single particle like interpretation,
indeed it does not even act in a Fock
space as it is a coherent state operator.
The perturbative expression of this is the infra-red problem in
QED: one cannot extract poles from the on-shell Green's functions of the theory
and hence an $S$-matrix cannot be constructed \cite{Zwanziger:1975jz,Papanicolaou:1976sv}.
Such observations have led to the view  among theorists \cite{kulish:1970}  that
no particle description is possible for the electron. Our point of view on this is that the
matter field is simply not
physical and we should not be shocked if its asymptotic limit is also
unphysical: it is far too early to give up on the electron as a particle!

In order to identify just what does describe the  physics in the asymptotic regime we
minimally need to
find fields which commute with the $B$ field. To this end we observe that at early and
late times the $B$ field can be written solely in terms of the free
electro-magnetic potential:
\begin{equation}\label{2bas}
  B(x)\to-\pa^\mu\Aas_\mu(x)=-\pa^\mu\Afree_\mu(x)\,.
\end{equation}
This result follows from (\ref{2aas}), and the conservation of the asymptotic
current if
\begin{equation}\label{2newcon}
  \lim_{a\to\infty}\inty\frac1{|y|}\delta(a-|y|)J_0^\as(a,y)=0\,.
\end{equation}
To demonstrate this, we insert the explicit expression (\ref{2jas}) for the
asymptotic current into (\ref{2newcon}). This generates a product of delta functions whose
arguments, it can be immediately seen, cannot both vanish for massive matter fields.

Hence the
modes of the $B$-field responsible for identifying the physical states
(\ref{2phys}) are asymptotically given by the modes of the free vector
potential as
\begin{equation}\label{2bisa}
 c(k)=ik\cd a(k)\,.
\end{equation}
Thus, the modes of the free fermionic field (\ref{2psifree}) are  physical in the asymptotic regime.
Hence  $\dd(q,s)\ket0$ \emph{is} indeed an asymptotic, charged one-particle state.
But it is  \emph{not} the  asymptotic limit of the Lagrangian matter
field. Given that this  charged state is created by a free creation operator,
one might now ask if it indeed has an  associated  electromagnetic
field. The important point to note is that the asymptotic
potential (\ref{2aas}) is \emph{not} that of a free theory. In
particular,
using (\ref{2d}) and
(\ref{2aas}), we see that
\begin{equation}\label{2abdag}
  [\Aas_\mu(x),\dd(q,s)]=-\frac{e}{4\pi}\frac{q_\mu}{\sqrt{(q\cd
  x)^2-q^2x^2}}\dd(q,s)\,.
\end{equation}
We recall from Chap.~14 of \cite{jackson:1998}
that the classical electro-magnetic
potential associated with a charge moving with momentum $q$
is, at large distances, given by
\begin{equation}\label{2cla}
 A_\mu^{\mathrm{class}}(x)=-\frac{e}{4\pi}\frac{q_\mu}{\sqrt{(q\cd
  x)^2-q^2x^2}}\,.
\end{equation}
Thus the  electromagnetic potential associated with the state
$\dd(q,s)\ket0$ indeed corresponds, at large distances, to that of a
particle moving with momentum $q$.
Hence  the asymptotically physical operator  $\dd(q,s)$ creates both a charged particle state
with momentum $q$ and its associated electromagnetic potential.

To summarise: we have seen that $\dd(q,s)$ is physical and creates
the correct electromagnetic field expected for a moving charge. It
is, however, \emph{not} directly related to the the asymptotic
limit of the matter field $\psi$. Is there then a (gauge
invariant) operator which has $\dd(q,s)$ as its asymptotic limit?
This is an important question since if there is such a field, then this would
correspond asymptotically  to an electron. The construction of such a field will be the
subject of the next section.

\section{Static Charges}

We have seen  that it is indeed possible to construct gauge
invariant, charged particle states at asymptotic times in QED. In fact, these
states are elements of the familiar Fock space associated with the modes of the
free fermionic field. In this sense, QED is no different from any other theory
with purely massive fields.  Where QED differs from other (non-gauge) theories
is in the fact that these fields  are \emph{not} the asymptotic limits of
the matter fields  from which the Lagrangian of the
theory was constructed. The neglect of this simple fact generates the infra-red
problem, prevents the construction of $S$-matrix elements and leads to the
abandonment of a particle description of charges. In order to motivate  our
general construction of the fields which \emph{do} asymptotically correspond to
the charged particle states, we will, in this section, reanalyse Dirac's
construction of the static charged field  and show in what sense it
correctly describes a static charged particle.

Dirac's construction of the field $\psidirac$ raises many fundamental questions
that we need to address if we are to make progress in our aim to describe
charged fields in both QED and, more generally, in the standard model. For
example, it is far from clear how unique this construction is or how it should
be extended to a non-abelian theory such as QCD where the chromo-electric field
of a static charge is not known \emph{a priori}.

Before analysing the uniqueness of the construction, let us first of all
see how a static interpretation of $\psidirac$ emerges even when the fields are not
infinitely massive. To this end we  define the gauge invariant  annihilation operator
$b(q,s,t,0)$ by
\begin{equation}\label{3bo}
  b(q,s,t,0):=\intx\frac1{\sqrt{2E_{\smash{q}}}}
  u^{\dag s}(q)\exp\bigg(-ie\frac{\pa_iA_i}{\nabla^2}\bigg)(x)\psi(x)\ee^{iq\ecd
  x}\,.
\end{equation}
The momentum $q$ is taken to be on-shell in this definition.
The additional label \lq0\rq\ in this operator reflects the conjectured
static nature of Dirac's construction.

At $O(e)$,  the proposed annihilation operator (\ref{3bo}) is the sum of two
terms, one from the matter as seen in Sect.~2 and one coming from the expansion of the dressing:
\begin{equation}\label{3boe}
  b^{[1]}(q,s,t,0)=\intx\frac1{\sqrt{2E_{\smash{q}}}}
  u^{\dag s}(q)\psi(x)\ee^{iq\ecd
  x}-ie\intx\frac1{\sqrt{2E_{\smash{q}}}}
  u^{\dag s}(q)\frac{\pa_iA_i}{\nabla^2}(x)\psi(x)\ee^{iq\ecd
  x}\,,
\end{equation}
where the superscript signifies that we only retain terms up to order $e$.
Starting in the interaction picture, then transforming to the
asymptotic Heisenberg picture will result in the first term in
(\ref{3boe}) becoming the $O(e)$ distorted annihilation operator
(\ref{2bbadr}). Given that the second term in (\ref{3boe}) is
already at $O(e)$, its asymptotic limit in this approximation will simply be the large
time limit of the operator expressed in terms of the  free fields
(\ref{2psifree}) and (\ref{2afree}). Using the identity
\begin{equation}\label{3naba}
  \frac{\pa_iA^\free_i}{\nabla^2}(x)=i\intk\frac1{2\omega_k}\bigg(
  \frac{\kb\cd\ab(k)}{\kb^2}\ee^{-ik\ecd x}-\frac{\kb\cd\ab^\dag
  (k)}{\kb^2}\ee^{ik\ecd x}\bigg)
  \,,
\end{equation}
and the Kulish-Faddeev argument discussed in Sect.~2, the second
term in (\ref{3boe}) can be shown to have  the asymptotic form
\begin{equation}\label{3baoe}
  e b(q,s)\intks\frac1{2\omega_k}\bigg(
  \frac{\kb\cd\ab(k)}{\kb^2}\ee^{-itk\ecd q/E_q}-\frac{\kb\cd
  \ab^\dag(k)}{\kb^2}\ee^{itk\ecd q/E_q}\bigg)
  \,.
\end{equation}
Combining the two contributions to $b^{[1]}(q,s,t,0)$, we get, in
the large time limit, the modified distorted annihilation operator
\begin{eqnarray}
 b^{[1]}(q,s,t,0)\to b(q,s)\Bigg\{1+&&\!\!\!\!\!\!\!\!e\intks\frac1{2\omega_k}
 \bigg[
  \bigg(\frac{\kb\cd\ab(k)}{\kb^2}-\frac{q\cd a(k)}{q\cd k}\bigg)
  \ee^{-itk\ecd q/E_q}\nonumber\\
  &&\qquad-\bigg(\frac{\kb\cd\ab^\dag(k)}{\kb^2}-\frac{q\cd a^\dag(k)}{q\cd k}
  \bigg)\ee^{itk\ecd
  q/E_q}\bigg]\Bigg\}
  \,.
\end{eqnarray}
In general this distortion does not vanish and thus, even though we are
dealing with a gauge invariant field, it does not  allow for a
charged particle interpretation. However, if the 4-momentum $q$ is at
the static point in the mass-shell, i.e., when  the four-momentum $q^\mu$ is
$m\eta^\mu$, where $\eta$ is the unit temporal vector,
then the distortion terms are
\begin{eqnarray}\label{3baoeall}
-e\intks\frac1{2\omega_k}
 \bigg(
  \frac{k\cd a(k)}{\kb^2}\ee^{-it\omega_k}
  \!\!\!\!\!\!\!\!\!&&-\frac{k\cd a^\dag(k)}{\kb^2}
  \ee^{it\omega_k}\bigg)\nonumber\\
  &&=
e\intks\frac{i}{2\omega^3_k}
 (
  c(k)\ee^{-it\omega_k}
  +c^\dag(k)
  \ee^{it\omega_k})\,,
\end{eqnarray}
where we have used the asymptotic identification (\ref{2bisa}) of the modes of
the $B$ field. As we have seen, these modes commute with the free annihilation
operator $b(q,s)$ and with themselves. Hence, using the definition of physical
states, (\ref{2phys}), we see that
between physical states the distortion operator reduces to the identity operator
at the static point in the mass shell.

The above analysis  shows that  matter dressed \`a la
Dirac,  $\psidirac$, has a particle interpretation. This interpretation we
stress
\emph{only holds at the static point in the mass shell}. Before discussing the exponentiation of
this result, or indeed its extension to moving charges, we need to understand
the uniqueness of the construction. To this end, we  need to step back
and ask where this solution came from, i.e., how should we characterise the
dressing so that it describes a charged particle with a well defined velocity?

Let us  initially look at the simpler situation where we have a heavy
matter field, $\varphi(x)$. This field can be thought of as  the infinite
mass limit of either a scalar or Dirac field which is, for the moment, not
coupled to a gauge theory. In this limit velocity is superselected \cite{Georgi:1990um} and the
equation of motion for the field has the universal form
\begin{equation}\label{4heavy}
 u\cd\pa\varphi(x)=0\,,
\end{equation}
where $u$ is the associated 4-velocity of the heavy particle.
This equation is simply the statement that the field is constant along the
world line of a particle moving with 4-velocity $u$.
Indeed, if we parameterise the world line of a particle moving with 4-velocity
$u^\mu=\gamma(\eta+v)^\mu$ (where $\eta$ is the unit time-like vector introduced
above, $v=(0,\vb)$ is the velocity and $\gamma=(1-|\vb|^2)^{-1/2}$) as
\begin{equation}\label{4parwl}
  x^\mu(s)=x^\mu+(s-x^0)(\eta+v)^\mu\,,
\end{equation}
then (\ref{4heavy}) implies that for arbitrary $s$,
\begin{equation}\label{4masssol}
  \varphi(x(s))=\varphi(x(0))\,.
\end{equation}

If the heavy matter field is now minimally coupled then the equation of motion
(\ref{4heavy}) becomes
\begin{equation}\label{4hmg}
  u\cd D\varphi(x)=0\,,
\end{equation}
where $D_\mu=\pa_\mu-ieA_\mu$. There is now no heavy charged particle interpretation to this equation. However
 the field $\varphi$ should not, and
indeed
cannot, be identified as a physical field since it is not gauge invariant. The
lack of a simple particle interpretation to the field $\varphi$ is thus not a
serious problem since what we need to do first is to dress this field. We will
demand that the dressing is such that, in addition to restoring gauge
invariance, it should also ensure that a (heavy) particle interpretation for the
dressed matter field holds.

We define the gauge invariant charged matter field by
\begin{equation}\label{4hdm}
 \Phi(x)=h^{-1}(x)\varphi(x)\,,
\end{equation}
where the dressing $h^{-1}(x)$ is such that under a gauge transformation we
have
\begin{equation}\label{3dgf}
 h^{-1}(x)\to h^{-1}(x)\ee^{-ie\theta(x)}\,.
\end{equation}
That is, a physical particle corresponds to the heavy matter field dressed with some
\lq brown muck\rq, whose exact form we will now clarify.
For this field to correspond to a heavy charged particle we must have
$u\cd\pa\Phi=0$. This follows from (\ref{4hmg}) if the dressing satisfies
\begin{equation}\label{4de}
  u\cd\pa h^{-1}(x)=-ieh^{-1}(x)u\cd A(x)\,.
\end{equation}
We call this equation the \emph{dressing equation.}
Equations (\ref{3dgf}) and (\ref{4de}) are the fundamental requirements on any
description of charged particles and are central to what follows.

Although our motivation for
this equation emerged from an analysis of the heavy matter sector, we demand
that, more generally, it holds  for the dressing of any field that is asymptotically
corresponding to a charged particle moving
with velocity~$u$. There are two arguments for this: it is well known that the
asymptotic dynamics of QED is governed by soft photons for whom any electron is
heavy; secondly, it can be shown\cite{Horan:1998im} that the asymptotic interaction Hamiltonian
vanishes for the propagator of the dressed fields which satisfy  (\ref{4de}) if one is at the right point
in the mass-shell.
We now specialise to the static situation and leave the
general case to the next section.

In the static situation the dressing equation (\ref{4de}) becomes
\begin{equation}\label{4des}
  \pa_0 h^{-1}(x)=-ieh^{-1}(x) A_0(x)\,.
\end{equation}
The first thing to note about this is that Dirac's proposal (\ref{diracelec}) \emph{does not
satisfy} this equation! Indeed, using the identity
\begin{equation}\label{4derexp}
 \pa_\mu \ee^O=\ee^O\big(\pa_\mu O+\tfrac12[\pa_\mu O,O]\big)
\end{equation}
where $O$ is an arbitrary operator whose commutator $[\pa_\mu O,O]$ is a
c-number, we find that
\begin{equation}\label{4staticprob}
\pa_0\exp\bigg(-ie\frac{\pa_iA_i}{\nabla^2}(x)\bigg)
=-ie\exp\bigg(-ie\frac{\pa_iA_i}{\nabla^2}(x)\bigg)\Bigg(
\frac{\pa_0\pa_jA_j}{\nabla^2}(x)
+\frac{e}2\intk\frac1{\omega_k^2}
\Bigg)\,,
\end{equation}
and even if we ignore the field independent term in the last factor, we do not
solve (\ref{4des}).

To  understand the relationship between Dirac's dressing and the static version
of the dressing equation we need to solve equation (\ref{4des}).
As is well known, the solution to equations like (\ref{4des}) have the form of an anti-time
ordered exponential:
\begin{equation}\label{3antt}
  h^{-1}(x,a)\approx
  \widetilde{T}\exp\bigg(-ie\int_a^t A_0(s,\xb)\,ds\bigg)\,,
\end{equation}
where $a$ is, as yet, undetermined.
Although this clearly solves (\ref{4des}), it does \emph{not} satisfy the gauge
transformation property (\ref{3dgf}) which is essential for a dressing. Indeed,
under a gauge transformation we have
\begin{equation}\label{3anttgt}
  h^{-1}(x,a)\to \ee^{ie\theta(a,\xb)}h^{-1}(x,a)\ee^{-ie\theta(x)}\,.
\end{equation}
There is no choice for the parameter $a$ for which $\theta(a,\xb)$ vanishes
(the choice $a=\pm\infty$ evaluates $\theta$ at time-like infinity where there
are no restrictions on the gauge transformations, see Fig.~\ref{prdg}).

We can  compensate for this bad behaviour under gauge transformations, while
still solving the dressing equation (\ref{4des}), by taking instead
\begin{equation}\label{3anttplus}
  h^{-1}(x,a)=\exp\bigg(-ie\frac{\pa_iA_i}{\nabla^2}(a,\xb)\bigg)\widetilde{T}
  \exp\bigg(-ie\int_a^t A_0(s,\xb)\,ds\bigg)\,.
\end{equation}
This we can write  as
\begin{equation}\label{3anttplus2}
  h^{-1}(x,a)=\widetilde{T}
  \exp\bigg(-ie\int_a^t \Big\{A_0(s,\xb)-\frac{\pa_0\pa_iA_i}{\nabla^2}(s,\xb)\Big\}\,ds\bigg)
  \exp\bigg(-ie\frac{\pa_iA_i}{\nabla^2}(x)\bigg)\,,
\end{equation}
where we have combined the two exponentials of (\ref{3anttplus})
under one exponential, then written one part as a total derivative. We have  neglected, for the moment, possible
commutator terms  which will be discussed
in Sect.~4.
The first term in this expression is now gauge invariant, a fact which
becomes manifest when we write this solution as
\begin{equation}\label{3anttplusf}
  h^{-1}(x,a)=\widetilde{T}
  \exp\bigg(ie\int_a^t \frac{\pa^\mu F_{\mu0}}{\nabla^2}(s,\xb)\,ds\bigg)
  \exp\bigg(-ie\frac{\pa_iA_i}{\nabla^2}(x)\bigg)\,.
\end{equation}
Thus we see a factorisation of the static dressing into a \emph{minimal part}
 which is essential for gauge invariance (and is just Dirac's original proposal for the static
dressing), plus an \emph{additional part} which is separately gauge invariant.

This general solution to the static dressing equation takes on a much simpler form
in the asymptotic regime where a charged particle
picture should emerge. Using the asymptotic space-time
commutators (\ref{2afcomm}) we see that
\begin{equation}\label{3fcomm}
 [\pa^\mu F^\as_{\mu\nu}(x), F^\as_{\lambda\rho}(y)]=0\,.
\end{equation}
This commutator implies two important results. Firstly, the additional part of
the dressing will commute with the electric and magnetic field operators, and
thus the electromagnetic field associated with the static dressing is the same
as that produced by the minimal, Dirac component. Hence we see that Dirac's
argument for the form of the dressing was not sensitive enough to detect the
additional term. Secondly,  we can use this commutator to dispense with the anti-time ordering in the dressing
(\ref{3anttplusf}) so that  in the asymptotic regime
\begin{equation}\label{3hasstatic}
  h^{-1}(x,a)\to
  \exp\bigg(ie\int_a^t \frac{\pa^\mu F^\as_{\mu0}}{\nabla^2}(s,\xb)\,ds\bigg)
  \exp\bigg(-ie\frac{\pa_iA^\as_i}{\nabla^2}(x)\bigg)\,.
\end{equation}
Using  Gauss' law (\ref{gausslaw}) in the asymptotic domain,  and the identification (\ref{2aas}), we
see that the static dressing factorises in the asymptotic regime into the
product of two terms.  One can be made out of the various matter contributions
and  is gauge invariant, it is analogous to the phase term in the distortion operator for the matter
field. The other part of the dressing  is constructed out of the free vector potential and, for
reasons that will become immediately apparent, we  call it the soft part of
the dressing:
\begin{equation}\label{3hsoft}
 h^{-1}_\soft(x)=\exp\bigg(-ie\frac{\pa_iA^\free_i}{\nabla^2}(x)\bigg)\,.
\end{equation}
This is the most important component in QED and, as was shown in
\cite{Lavelle:1998dv}, its extension to QCD is the dominant glue
around quarks at short distances.

Armed with this result, we can now study the exponentiation of the soft contributions to the
asymptotic annihilation operator (\ref{3bo}). Repeating the Kulish-Faddeev
argument for the soft components of both the matter and dressing we find that
\begin{equation}\label{3softannihilation}
  b(q,s,t,0)\to h^{-1}_\soft(q,t)D_\soft(q,t)b(q,s)\,,
\end{equation}
where
\begin{equation}\label{3softfh}
   h^{-1}_{\soft}(q,t)=\exp\!\!\Bigg\{e\!\!\intks\frac1{2\omega_k}
  \bigg(
  \frac{\kb\cd \ab(k)}{\kb^2}\ee^{-itk\ecd q/E_q}-
\frac{\kb\cd \ab^\dag(k)}{\kb^2}\ee^{itk\ecd q/E_q}
  \bigg)\Bigg\}\,,
\end{equation}
which is the exponentiation of our previous result (\ref{3baoe}).
The combined effect of these distortion  factors can easily be evaluated using the
canonical commutation relations for the modes $a_\mu(k)$ and the
Baker-Campbell-Hausdorff  formula. One finds
\begin{eqnarray}
 h^{-1}_\soft(q,t)D_{\soft}(q,t)&=&\exp\Bigg\{e\!\!\intks\frac1{2\omega_k}
  \bigg[\bigg(
  \frac{\kb\cd \ab(k)}{\kb^2}-\frac{q\cd a(k)}{q\cd k}\bigg)\ee^{-itk\ecd q/E_q}\nonumber\\
  &&\qquad\qquad\qquad-
\bigg(\frac{\kb\cd \ab^\dag(k)}{\kb^2}-\frac{q\cd \ad(k)}{q\cd k}\bigg)\ee^{itk\ecd q/E_q}
  \bigg]\Bigg\}\,.
\end{eqnarray}
Note that the commutator in the BCH-formula vanishes here.
Just as we saw in (\ref{3baoeall}), this becomes the identity operator between physical
states at the correct point in the mass-shell where the momentum $q^\mu$ is
at the static point in the mass shell: $q^\mu=m\eta^\mu$.

To summarise: we have seen in this section how to characterise the dressing appropriate
to a charged particle  moving with a given velocity. In the static case we solved the dressing
equation and saw how Dirac's proposal for a static charge emerged as the
minimal part of the static dressing. Using this construction we showed how the static annihilation
operator  emerges as the asymptotic limit of the statically dressed matter field.
Having seen how to describe static charges, we now proceed to an analysis of  moving charges.

\section{Moving Charges}

In this section we will show that the dressing needed to describe a charged particle
moving with four-velocity $u^\mu$ is
\begin{equation}\label{4sol}
  h^{-1}(x)=\ee^{-ieK(x)}\ee^{-ie\chi(x)}\,,
\end{equation}
where the minimal part of the dressing is  given by
\begin{equation}\label{4min}
  \chi(x)=\frac{\G\cd A}{\G\cd\pa}\,,
\end{equation}
with $\G^\mu=(\eta+v)^\mu(\eta-v)\cd\pa-\pa^\mu$,
and the additional (gauge invariant) part of the dressing is
\begin{equation}\label{4add}
  K(x)=\int_{\pm\infty}^{x^0}(\eta+v)^\mu\frac{\pa^\nu
  F_{\nu\mu}}{\G\cd\pa}(x(s))\,ds\,.
\end{equation}
In this last expression the integral is along the world-line of a massive particle with
four-velocity $u^\mu$ parameterised as in (\ref{4parwl}).
Before deriving this result from the dressing equation (\ref{4de}), we will first
verify that this dressing generates the correct electric and magnetic fields and
demonstrate that its modes have the asymptotic
limit appropriate to a charged particle moving with velocity $\vb$.

In the asymptotic regime we can again argue from (\ref{3fcomm}) that the
additional component of the dressing does not affect the electromagnetic
configuration. This allows us to define the minimally dressed charged field $\psi_v$
by
\begin{equation}\label{3psisubv}
  \psi_v(x)=\ee^{-ie\chi(x)}\psi(x)\,.
\end{equation}
By construction, this is gauge invariant. It is also  straightforward to see that this field is physical in
the context of the $B$-field formalism discussed in Sect.\ 2. In particular,
from (\ref{2ca}) and (\ref{2cp}), it is easy to see that
\begin{equation}\label{3cphiv}
  [c(k),\psi_v(x)]=0\,.
\end{equation}
In the static limit this dressed field reduces to  Dirac's
expression (\ref{diracelec}), i.e.,  $\psi_0=\psidirac$.

Repeating the argument given by Dirac leading up to the  expression (\ref{stt})
for the electric field of the static charge, the electromagnetic fields
generated by (\ref{3psisubv}) are given by the equal-time commutators
$-ie[E_i(x),\chi(y)]_{_{\mathrm{et}}}$ and
$-ie[B_i(x),\chi(y)]_{_{\mathrm{et}}}$.
In order to evaluate these  we first  need
to make clear how $1/\G\cd\pa$ is defined in (\ref{4min}).

We take
\begin{equation}\label{4def1gd}
  \frac1{\G\cd\pa}f(\xb):=\intz G(\xb-\zb)f(\zb)\,,
\end{equation}
where
\begin{equation}\label{4gfn}
  \G\cd\pa G(\xb-\zb)=\de^3(\xb-\zb)
\end{equation}
and we note from its definition that $\G\cd\pa=\nabla^2-(\vb\cd\pa)^2$. Taking
the Fourier transform of (\ref{4gfn}) we see that
\begin{equation}\label{4gvft}
G(\xb)
=-\intk\frac1{\kb^2-(\vb\cd\kb)^2}e^{i\kb\ecd \xb}\,.
\end{equation}
This integral is   computed in the standard fashion by first  diagonalising
the quadratic form in the denominator. The final expression for $G(\xb)$ is
then
\begin{equation}\label{3gvft}
G(\xb)=-\frac1{4\pi}\frac{\gamma}{\sqrt{\xb^2+\gamma^2(\vb\cd\xb)^2}}\,.
\end{equation}
Note that in the static limit this
reproduces  (\ref{sttt}).

The electric and magnetic fields generated by this dressing are now straightforward to
calculate. For example, applying the operator (\ref{3psisubv}) to a state $\ket0$ with no electric field, we find
\begin{equation}\label{4etev}
 E_i(x)\psi_v(y)\ket0=-\frac{e}{4\pi}
  \frac{\gamma(\xb-\yb)_i}{\big(|\xb-\yb|^2+\gamma^2(\vb\cdot(\xb-\yb))^2\big)^{\frac32}}\psi_v(y)\ket0\,.
\end{equation}
This is precisely the electric field associated with a charged particle moving with
velocity~$\vb$.

As we saw with the static  charge, the interpretation of the
operator $\psi_v(x)$ as that creating a moving charge at the point $x$ only
holds in the infinite mass limit. More generally then, we extend  the definition of the static, gauge invariant
annihilation operator (\ref{3bo}) to the moving case  by defining
the mode
\begin{equation}\label{3bv}
  b(q,s,t,v)=\intx\frac1{\sqrt{2E_{\smash{q}}}}
  u^{\dag s}(q)\exp\bigg(-ie\frac{\G\cd A}{\G\cd\pa}\bigg)(x)\psi(x)\ee^{iq\ecd
  x}\,.
\end{equation}
We will now  demonstrate that this is  a particle  annihilation operator at the right point in the
mass-shell.

At $O(e)$,  $b^{[1]}(q,s,t,v)$  is the sum of two terms in which the free field expansions
for the potential in the dressing can be used. Using the expression
\begin{equation}\label{3nabav}
  \frac{\G\cd A^\free}{\G\cd\pa}(x)=i\intk\frac1{2\omega_k}\bigg(
  \frac{V\cd a(k)}{V\cd k}\ee^{-ik\ecd x}-\frac{V\cd a^\dag
  (k)}{V\cd k}\ee^{ik\ecd x}\bigg)
  \,,
\end{equation}
where we have introduced the notation that
\begin{equation}\label{3vdef}
  V^\mu=(\eta+v)^\mu(\eta-v)\cd k-k^\mu\,,
\end{equation}
we see that in the large time limit
\begin{eqnarray}
 b^{[1]}(q,s,t,v)\to b(q,s)\Bigg\{1+&&\!\!\!\!\!\!\!\!e\intk\frac1{2\omega_k}
 \bigg[
  \bigg(\frac{V\cd a(k)}{V\cd k}-\frac{q\cd a(k)}{q\cd k}\bigg)
  \ee^{-itk\ecd q/E_q}\nonumber\\
  &&\qquad-\bigg(\frac{V\cd a^\dag(k)}{V\cd k}-\frac{q\cd a^\dag(k)}{q\cd k}
  \bigg)\ee^{itk\ecd
  q/E_q}\bigg]\Bigg\}
  \,.
\end{eqnarray}
In the distortion term we note that we can write (recall that $k$ is on-shell)
\begin{eqnarray}
\frac{V^\mu}{V\cd k}-\frac{q^\mu}{q\cd k}&=& \frac{(\eta+v)^\mu(\eta-v)\cd
k-k^\mu}{(\eta+v)\cd k(\eta-v)\cd k}-\frac{q^\mu}{q\cd k}\nonumber\\
&=&\frac{(\eta+v)^\mu}{(\eta+v)\cd k}-\frac{q^\mu}{q\cd k}-\frac{k^\mu}{V\cd k}\,.
\end{eqnarray}
Hence, at the point in the mass-shell where $q^\mu=m\gamma(\eta+v)^\mu$, the
distortion operator becomes the trivial operator
\begin{eqnarray}
-e\intk\frac1{2\omega_k}
 \bigg(
  \frac{k\cd a(k)}{V\cd k}\ee^{-it\omega_k}
  \!\!\!\!\!\!\!\!\!&&-\frac{k\cd a^\dag(k)}{V\cd k}
  \ee^{it\omega_k}\bigg)\nonumber\\
  &&=
e\intk\frac{i}{2\omega_k V\cd k}
 (
  c(k)\ee^{-it\omega_k}
  +c^\dag(k)
  \ee^{it\omega_k})\,,
\end{eqnarray}
which vanishes between physical states.

The exponentiation of this result for the soft  contributions to both the distortion operator and the
dressing,
follows in much the same way as it did for the static case.  Indeed, at large
times we get
\begin{equation}\label{3bexp}
  b(q,s,t,v)\to h_\soft(q,t,v)D_\soft(q,t)b(q,s)\,,
\end{equation}
where $D_\soft(q,t)$ is given in (\ref{2dsoft}) and
\begin{equation}\label{3ddress}
  h_{\soft}(q,t,v)=\exp\left\{e\!\!\!\intks\frac1{2\omega_k}
  \left(
  \frac{V\cd a(k)}{V\cd k}\ee^{-itk\ecd q/E_q}-
\frac{V\cd \ad(k)}{V\cd k}\ee^{itk\ecd q/E_q}
  \right)\right\}\,.
\end{equation}
The combined effect of these distortions yields the overall
soft
distortion of the particle mode as
\begin{eqnarray}
 h_\soft(q,t,v)D_{\soft}(q,t)&=&\exp\Bigg(e\!\!\intks\frac1{2\omega_k}
  \bigg[\bigg(
  \frac{V\cd a(k)}{V\cd k}-\frac{q\cd a(k)}{q\cd k}\bigg)\ee^{-itk\ecd q/E_q}\nonumber\\
  &&\qquad\qquad\qquad-
\bigg(\frac{V\cd \ad(k)}{V\cd k}-\frac{q\cd \ad(k)}{q\cd k}\bigg)\ee^{itk\ecd q/E_q}
  \bigg]\Bigg)\,.
\end{eqnarray}
Just as we saw above, this becomes the identity operator between physical
states at that point in the mass-shell where  $q^\mu=m\gamma(\eta+v)^\mu$.

This analysis has shown that the dressed matter $\psi_v$ can be interpreted as
a charged particle moving with velocity $v$  at the appropriate  point in the
mass shell. In this sense, we see that a relativistic description of charged
particle is indeed possible. This dressing (\ref{4sol})  for a moving charge
can, of course, be derived from the dressing
equation (\ref{4de}). Full details of this are given in the appendix.


\section{From Dressed Matter to the $S$-Matrix}\label{fifth}

In this section we want to show how the $S$-matrix elements of QED
can be extracted in terms of Green's functions of dressed matter
fields. We will see that the usual perturbative techniques can be
applied. It should become clear that a correct application of the usual
LSZ~formalism (for a discussion with fermionic fields
see, e.g., Chap.~5 of Ref.~\cite{kaku:1993})
requires the use of dressed matter.

\begin{figure}
\begin{center}
$$
%
\includegraphics{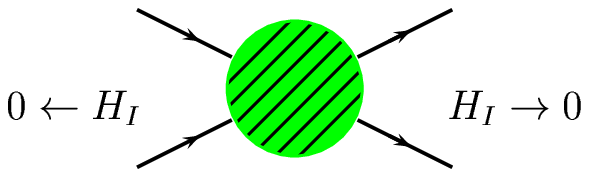}
$$
\end{center}
        \caption{Sketch of a scattering process where the coupling is assumed to vanish
        for large times.}
        \label{gvanish}
\end{figure}

Recall that in the usual LSZ~route to the $S$-matrix it is assumed
that there is an overlap at large times between the interacting
fields and the free fields. This is schematically illustrated in
Fig.~\ref{gvanish}. Since we have seen that this overlap does
not take place between the usual matter fields in the QED Lagrangian,
it is clear that we should expect problems in implementing the LSZ
procedure and, indeed, one finds that the usual Green's functions
do not have a good pole structure. The responses to this range
from using coherent states and abandoning the pole structure~\cite{schroer:1963})
to giving up the idea of $S$-matrix elements and only considering
inclusive cross-sections including all possible final state photons.
We now want to show how the use of dressed fields in QED lets us
follow the LSZ path to the $S$-matrix. The companion paper, \II,
will show in perturbative studies that good pole structures are
obtained with the use of these fields.

Consider the simplest case: one particle goes to one particle.
The equivalent $S$-matrix element may be expressed as
\begin{equation}
\bra{\hbox{out}}
{\hbox{in}}\rangle
=
\bra{\hbox{out}}
b^{\dag}(q)
\ket{0}
  \,,
\end{equation}
where $\ket{0}$ is the vacuum state and $b^{\dag}$ ($b$) are just the
traditional one-particle Fock creation (annihilation) operators.
In the free theory we may express the creation operator by
\begin{equation}\label{creator}
b^{\dag}(q)=\intx \normm{q} \psi^{\dag}(x) u(q) \hbox{e}^{-iq\cdot x}
\,.
\end{equation}
Now it is normally \emph{assumed} that for large times the usual
Heisenberg fields approach the free fields. However, we have seen
in Sect.~2 that this is not the case in unbroken gauge theories
like QED. Rather, as was demonstrated in  Sect.'s~3 and~4, we have at
each point, $q$, on the mass shell an appropriately dressed field,
$\psi_v$, such that  in (\ref{creator})
we may, if we neglect the unobservable phase effects, (weakly)
replace $\psi_v\to\psi^f$ for $t\to\pm\infty$. (Note that for simplicity
we do not explicitly insert renormalisation constants. Renormalisation
will be studied in the explicit calculations of \II.)

We thus see that in QED we obtain
\begin{equation}
  \bra{\hbox{out}}
b^{\dag}(q)
\ket{0}=
\lim_{t\to-\infty} \intx
\normm{q}   \bra{\hbox{out}}
\psi^{\dag}_v(x) \ket{0}
u(q) \hbox{e}^{-iq\cdot x}
\,.
\end{equation}
In the standard fashion we may introduce an integration over time
to obtain
\begin{eqnarray}
  \bra{\hbox{out}}
b^{\dag}(q)
\ket{0}&=&
-\int\!d^4x\frac{d}{dt}
\normm{q}   \bra{\hbox{out}}
\psi^{\dag}_v(x) \ket{0}
u(q) \hbox{e}^{-iq\cdot x}\nonumber \\
&&\qquad +
\lim_{t\to\infty} \intx
\normm{q}   \bra{\hbox{out}}
\psi^{\dag}_v(x) \ket{0}
u(q) \hbox{e}^{-iq\cdot x}
\,.
\end{eqnarray}
As is usual this second, reduced term corresponds to disconnected
graphs and may be dropped.

The time derivative acts upon both the matter field and the
exponential. Writing $\psi^{\dag}_v=\bar\psi_v\gamma_0$ and using
that $(q\!\!\!\slash-m)u(q)=0$, we find
\begin{equation}
  \bra{\hbox{out}}
b^{\dag}(q)
\ket{0}=
i\int\!d^4x \bra{\hbox{out}}\bar \psi_v(x) \ket{0}
(i{\overleftarrow{\not\!\partial}}+m) \normm{q}u(q)\hbox{e}^{-iq\cdot x}
\,.
\end{equation}
We note that this is the standard result, but \emph{the matter
field is replaced by the appropriate dressed matter}.

We now want to continue by using an annihilation operator
to replace the out state by the vacuum. using the counterpart of
(\ref{creator}), and taking time-ordering into account, we rapidly find
\begin{eqnarray}\label{pdp}
\bra{\hbox{out}}
{\hbox{in}}\rangle
&=&
\bra{0}
b(p) b^{\dag}(q)
\ket{0}\nonumber\\
&=&
\int\!d^4x \int\!d^4 y
 (i{{\not\!  \partial}}_y-m)
\bra{0}T\left( \psi_{v'}(y) \bar \psi_v(x)\right) \ket{0}
(i{\overleftarrow{\not\!\partial}}_x+m) \nonumber\\
&& \qquad \qquad \qquad \quad \times \normm{q}u(q)\hbox{e}^{-iq\cdot x}
  \normm{p}u(p)\hbox{e}^{ip\cdot y}
  \,,
\end{eqnarray}
where $p$ (and $v'$) correspond to the momentum (velocity) of the
outgoing charge. Clearly this can be generalised to more
interesting $S$-matrix elements.

We now want to reexpress (\ref{pdp}) in terms of a Gell-Mann--Low
formula. The usual method relies upon the existence of a unitary
operation linking the Heisenberg and the free fields. We have already
seen in this paper that the free matter fields
are not unitarily related
to the equivalent Heisenberg fields, but  for
the dressed fields we indeed have
\begin{equation}\label{units}
\psi_v(t,\xb)=U^{-1}(t)\psi_v^{\rm f}(t,\xb)U(t)
  \,,
\end{equation}
where  $U$ is the usual operator, made up from exponentiating the
interaction Hamiltonian expressed in terms of free fields.
We stress that this relation only holds at the right point on the mass
shell appropriate to the dressing.

Once we have this relation it is straightforward to repeat the
usual procedure to obtain the standard Gell-Mann--Low formula but
with the charged matter fields replaced everywhere by appropriately
dressed matter. With this, we are in a position
to perform perturbative tests. These are presented in the companion paper, \II.


\section{Conclusions}

In this paper we have shown how  to describe
charged particles in relativistic QED through a process of dressing
the matter with the appropriate electromagnetic cloud. The essential ingredients
which we used to obtain the precise structure of the dressing
were  gauge invariance (\ref{3dgf}) and the kinematical
requirement which we call the dressing equation (\ref{4de}).

We solved these two conditions and obtained a structured dressing
composed of two factors: a minimal component which had the correct
gauge transformation properties and an additional, gauge invariant
part which was necessary to fulfill the dressing equation. In this
paper we have focused primarily on the minimal part of the
dressing and, in particular, we have shown that it removes the
soft distortion factor which has been previously claimed to
prevent a particle interpretation of QED. This has been
demonstrated for a charged particle moving with an arbitrary
velocity.

The vanishing of the distortion factor shows that the
long range interactions of QED, due to the masslessness of the photon, are
encoded in the dressing rather than any residual interaction term in the Hamiltonian.
An important consequence of this is that the asymptotic
Hamiltonian becomes the free Hamiltonian for our dressed fields.
This means that the LSZ-formalism can be directly carried out
without any use of a dubious adiabatic \lq switching off\rq\ of
the QED coupling.

The process of dressing a gauge dependent variable so as to make it gauge invariant
is a very general procedure. How one would want to dress the fields that form a
bound state might be very different from how individual charges should be
dressed. In this work we are only interested in dressing a charged field so
that it corresponds to a charged particle moving with the appropriate velocity
at either early or late times. The interesting topic of how to describe, say, a
bound state such as positronium, that is either entering or leaving a scattering
process, we leave to future work.

Turning to QCD, where we want to construct colour charged quarks
and gluons, we note first that our two fundamental requirements on
the dressing have immediate non-abelian extensions. A simple
algorithm for constructing the minimal, soft gluonic dressing
around a quark can be found in \cite{Lavelle:1997ty}. The
potential between two so-dressed quarks was studied in
\cite{Lavelle:1998dv}. There it was shown that this dressing generates the
anti-screening interaction of QCD which is responsible for
asymptotic freedom. We have thus identified the non-abelian extension of the soft
dressing as  the dominant glue around quarks. A corollary of this
result is that we see that the anti-screening interaction takes
place between two separately gauge invariant constituent quarks and there is as yet no sign of
any flux tube formation.

In the companion paper \II\ we will take the results of this paper
and submit them to a battery of tests. There we will see that the
particle of our dressed fields also emerges from perturbative
calculations. The Green's functions of these fields will be shown
to be free of on-shell infra-red divergences and to have the
pole structure expected of physical particles. Detailed studies of
the ultra-violet behaviour of these fields and their Green's
functions will be presented and it will be shown that the
renormalisation programme can be carried out without difficulties.

\no\textbf{Acknowledgements:} This work was supported by the British
Council/Spanish Education Ministry \textit{Acciones Integradas} grant
no.\  1801 /HB1997-0141. We thank Robin Horan,
Tom Steele, Shogo Tanimura and Izumi Tsutsui for discussions, the organisers of the
XIXth UK Theory Institute where some of
this work was carried out and PPARC for support from the Theory Travel Fund.
EB thanks the HEP group at BNL for their
warm hospitality and many interesting comments. He also acknowledges a
grant from the {\em Direcci\'on General de Ense\~nanza Superior e
Investigaci\'on Cient\'{\i}fica.}

\appendix

\section{Appendix}

In this appendix we show how the dressing for a moving charge (\ref{4sol})
follows directly from the gauge transformation properties (\ref{3dgf}) of the dressing and
the dressing equation (\ref{4de}).

Before presenting the solution, we need to briefly reexamine Dirac's
general formula (\ref{fdress}) for a gauge invariant field. In line
with (\ref{fdress}), we  take as an ansatz for the electro-magnetic cloud
an exponential of the form $\exp(-ie\chi(x))$
where, under a gauge transformation
\begin{equation}\label{3chitran}
  \chi(x)\to\chi(x)+\theta(x)\,.
\end{equation}
This is precisely the type  of transformation  that Dirac
investigated in (\ref{fdress}) and we would be tempted to accept as the
general solution
\begin{equation}\label{dansatz}
  \chi(x)=\int\!d^4z\,f^\mu(x-z) A_\mu(z)\,,
\end{equation}
where $f^\mu(x-z)$ satisfies $\pa_\mu f^\mu(x-z)=\de^4(x-z)$. But
this only implies (\ref{3chitran}) if \emph{no surface terms arise
when we integrate by parts}. The restriction on the local gauge
transformations to those that vanish at spatial infinity is quite
natural as finite energy restrictions impose a $1/r$ fall-off on
the potential. However, as discussed around Fig.~\ref{prdg}, no such restriction applies to the fields
at temporal infinity. Thus, to maintain gauge invariance,  the form of $f^\mu(x-z)$ must be such
that no  surface terms arise at large times.

As it stands, we can only infer from this that $f^0(x-z)$ should be
zero outside of some bounded region in the $z^0$-direction. To get
more from this, we follow Dirac's proposal for the static charge and take
\begin{equation}\label{3ansatz}
  \chi(x)=\int\!d^4z\,G(x-z)\G^\mu A_\mu(z)\,,
\end{equation}
where $\G^\mu$ is a first order, differential operator and
$\G\cd\pa\, G(x-z)=\de^4(x-z)$. In order to avoid the surface terms
that would obstruct the gauge transformation properties of the
dressing, we must have that the operator $\G\cd\pa$
\emph{ cannot involve any time
derivatives.} Given this restriction, we see that
$G(x-z)=\de(x^0-z^0)G(\xb-\zb)$ where
\begin{equation}\label{3gpaong}
  \G\cd\pa\, G(\xb-\zb)=\de^3(\xb-\zb)\,.
\end{equation}
We shall, for convenience, write
$\chi$ as
\begin{equation}\label{3chinotation}
  \chi(x)=\int\!d^3z\,G(\xb-\zb)\,\G^\mu A_\mu(x_0,\zb):=\frac{\G\cd
  A}{\G\cd\pa}(x)\,,
\end{equation}
which mirrors the notation adopted in the body of this paper (cf. Sect.~4). So far,
of course, $\G$ remains undefined here.

We now make a more general ansatz,  based upon our experience in the static case, i.e., that the dressing
factorises into the product of two terms:
\begin{equation}\label{4dressing}
h^{-1}(x)=e^{-ieK(x)}e^{-ie\chi(x)}\,,
\end{equation}
where $K$ is gauge invariant and $\chi\to\chi+\theta$ under the gauge
transformation (\ref{gauge}). The second term is thus a \emph{minimal part} of the
dressing, essential for the gauge transformation properties of the dressing.
The first term in (\ref{4dressing}) is then an \emph{additional part} of the
dressing. Here we write this as a simple exponential, rather than a
path-ordered exponential as we did for the static example in
(\ref{3antt}), since we are taking our fields to be the asymptotic
fields discussed in Sect.~2, and the commutators of these fields will be seen to
allow this ansatz. We note that \emph{all of the fields} in the rest of this
appendix are \emph{asymptotic} fields and we omit the \lq as\rq\
superscript in what follows.

For the
additional part of the dressing it is enough  in this abelian theory to also take $K$ to be  linear in the fields and
we find that it is sufficient  to make the expansion
\begin{equation}\label{4kine}
e K(x)=eK_1(x)+e^2K_2(x)\,.
\end{equation}
The dressing equation (\ref{4de}) can then be expanded in powers of the coupling to give the two equations:
\begin{equation}\label{4deq1}
(\eta+v)^\mu\pa_\mu(K_1+\chi)=(\eta+v)^\mu A_\mu\,,
\end{equation}
and
\begin{equation}\label{4deq2}
i(\eta+v)^\mu \pa_\mu
K_2=\tfrac12(\eta+v)^\mu\big(\pa_\mu[\chi,K_1]-[A_\mu,K_1+\chi]\big)\,.
\end{equation}
The solution to (\ref{4deq1}) is
\begin{equation}\label{4deq11}
  K_1(x)+\chi(x)=\int_a^{x^0}(\eta+v)^\mu A_\mu(x(s))\,ds+\chi(x(a))\,,
\end{equation}
where the integral is along the world line parameterised by (\ref{4parwl}) with starting point $a$.
The constant term
$\chi(x(a))$ is needed to ensure that $K$ is gauge invariant. This means that
we can write $K_1$ in the manifestly  invariant form:
\begin{eqnarray}\label{4k1}
K_1(x)&=&\int_a^{x^0}(\eta+v)^\mu\Big( A_\mu(x(s))-\pa_\mu\chi(x(s))\Big)\,ds\nonumber\\
&=&\int_a^{x^0}(\eta+v)^\mu\frac{\G^\nu F_{\nu\mu}}{\G\cd\pa}(x(s))\,ds\,.
\end{eqnarray}

The lower range of integration found in $K_1$ should, in this asymptotic
regime, have no physical significance. This will be ensured if the derivative
of the additional part of the dressing with respect to $a$ is zero between
physical states. This would be the case if this derivative either vanishes or is
constructed from the $B$-field.  From the form of (\ref{4k1}) we see that it will not vanish.
Therefore we must show that it is made from the $B$-field. We
thus postulate that it can be written as
\begin{eqnarray}\label{4a1}
\frac{\pa\ }{\pa a}K(x)&=&(\eta+v)^\mu\Big(\frac{\pa^\nu  F_{\nu\mu}}{\G\cd\pa}(x(a))
+e\frac{J_\mu}{\G\cd\pa}(x(a))\Big)\,,\nonumber\\
&=&(\eta+v)^\mu\frac{\pa_\mu B}{\G\cd\pa}(x(a))\,.
\end{eqnarray}
We now investigate what this implies for $K_1$ and $K_2$.

From equations (\ref{4k1}) and (\ref{4a1}) we see that
\begin{eqnarray}\label{4ka1}
\frac{\pa\ }{\pa a}K_1(x)&=&-(\eta+v)^\mu\frac{\G^\nu
F_{\nu\mu}}{\G\cd\pa}(x(a))\nonumber\\
&=&
(\eta+v)^\mu\frac{\pa^\nu  F_{\nu\mu}}{\G\cd\pa}(x(a))\,.
\end{eqnarray}
Hence, from the anti-symmetry of $F_{\nu\mu}$ and the fact that $\G^\nu$ is a
first order differential operator constructed out of $\pa^\nu$, $\eta^\nu$ and $v^\nu$,  we must have that
\begin{equation}\label{4gab}
  \G^\nu=-\pa^\nu+(\eta+v)^\nu(\alpha\eta\cd\pa+\beta v\cd\pa)\,.
\end{equation}
The constants $\alpha$ and $\beta$ can be fixed by the requirement that
$\G\cd\pa$ contains no time derivatives. It is easy to see that
\begin{equation}\label{4gabpa}
  \G\cd\pa=(\alpha-1)(\eta\cd\pa)^2+(\alpha+\beta)(\eta\cd\pa)(v\cd\pa)+
  \beta(v\cd\pa)^2+\nabla^2\,.
\end{equation}
Hence we must have that $\alpha=1$ and $\beta=-1$. From this we see that
\begin{equation}\label{4gfinal}
  \G^\nu=(\eta+v)^\nu(\eta- v)\cd\pa-\pa^\nu\,.
\end{equation}
Given this result we obtain
\begin{equation}\label{4k1final}
  K_1(x)=\int_a^{x^0}(\eta+v)^\mu\frac{\pa^\nu F_{\nu\mu}}{\G\cd\pa}(x(s))\,ds\,.
\end{equation}
Using the commutators (\ref{3fcomm}) it follows that
\begin{equation}\label{4kka}
  [K_1(x),\frac{\pa K_1}{\pa a}(x)]=0\,
\end{equation}
and hence the $a$-independence of the construction will follow from (\ref{4a1})
if
\begin{equation}\label{4ak2}
  \frac{\pa K_2}{\pa a}(x)=(\eta+v)^\nu\frac{J_\nu}{\G\cd\pa}(x(a))\,.
\end{equation}
This equation and (\ref{4deq2}) determine the form of $K_2$.

Given that the (asymptotic) current, $J$, is gauge invariant, we can
immediately solve (\ref{4ak2}) to get
\begin{equation}\label{4k2abit}
  K_2(x)=\int_{-\infty}^{a}(\eta+v)^\nu\frac{J_\nu}{\G\cd\pa}(x(s))\,ds+\mbox{$a$-independent terms.}
\end{equation}
Putting this expression into (\ref{4deq2}) we see that the
$a$-dependence cancels on the left-hand side. This means it cannot survive the commutators
in (\ref{4deq2}). This cancellation is far from obvious. To show that it does
happen we note that
{\setlength\arraycolsep{1pt}
\begin{eqnarray}\label{4k2firsta}
(\eta{+}v)^\mu[A_\mu(x),K_1(x){+}\chi(x)]&=&i\gamma^{-2}\!\!\intk
\frac1{\omega^2_k{-}(\kb\cd\vb)^2}\nonumber\\
&&-i\intk\frac1{\omega^2_k{-}(\kb\cd\vb)^2}
\ee^{i\kb\ecd\vb(x^0{-}a)}\cos\big(\omega_k(x^0{-}a)\big)\\
&&-i\intk\frac{i\kb\cd\vb}{\omega_k(\omega^2_k{-}(\kb\cd\vb)^2)}
\ee^{i\kb\ecd\vb(x^0{-}a)}\sin\big(\omega_k(x^0{-}a)\big)\,,\nonumber
\end{eqnarray}
}
and
{\setlength\arraycolsep{1pt}
\begin{eqnarray}\label{4k2seca}
[\chi(x),K_1(x)]
&=&-\int_{a}^{x^0}\!\!\!ds\intk\frac1{\omega^2_k{-}(\kb\cd\vb)^2}
\ee^{i\kb\ecd\vb(x^0{-}s)}\Big(\!\!\cos\big(\omega_k(x^0{-}s)\big)\nonumber\\
&&\hskip8cm+\sin\big(\omega_k(x^0{-}s)\big)\Big)\,.
\end{eqnarray}
}
Hence
{\setlength\arraycolsep{1pt}
\begin{eqnarray}\label{4k2thirda}
(\eta+v)^\mu\pa_\mu[\chi(x),K_1(x)]&=&\pa_0[\chi(x),K_1(x)]\\
&=&-i\intk\frac1{\omega^2_k{-}(\kb\cd\vb)^2}
\ee^{i\kb\ecd\vb(x^0{-}a)}\cos\big(\omega_k(x^0{-}a)\big)\\
&&-i\intk\frac{i\kb\cd\vb}{\omega_k(\omega^2_k{-}(\kb\cd\vb)^2)}
\ee^{i\kb\ecd\vb(x^0{-}a)}\sin\big(\omega_k(x^0{-}a)\big)\,,\nonumber
\end{eqnarray}
}%
which implies that that $a$-dependence has indeed been killed off.
Thus from (\ref{4deq2}) we see that
\begin{equation}\label{4k2end}
  i(\eta+v)^\mu\pa_\mu K_2(x)=-\tfrac12i\gamma^{-2}\!\!\intk
\frac1{\omega^2_k{-}(\kb\cd\vb)^2}\,.
\end{equation}
Combining this equation with (\ref{4ak2}) we can solve for $K_2$ to get
\begin{equation}\label{4k2sol}
K_2(x)=-\tfrac12\gamma^{-1}u\cd x\!\!\intk
\frac1{\omega^2_k{-}(\kb\cd\vb)^2}+\int_{-\infty}^{a}(\eta+v)^\nu\frac{J_\nu}{\G\cd\pa}(x(s))\,ds\,.
\end{equation}
This together with (\ref{4k1final}), (\ref{3chinotation}) and
(\ref{4gfinal}) is the general solution to the dressing equation and
the requirement of gauge invariance.
Since, by construction, this is independent of $a$, we can set
$a=-\infty$ and hence obtain the  result promised at the start of Sect.~4.

\end{document}